\documentclass[aps,prl,twocolumn,showpacs,superscriptaddress,showkeys]{revtex4-1}

\usepackage{amssymb}
\usepackage{amsmath}
\usepackage{amsfonts}
\usepackage{graphicx}
\usepackage{color}
\usepackage{xspace}
\usepackage{ulem}
\usepackage{mathtools}
\usepackage{hhline}

\newcommand{\AddrVandy}{Department of Physics and Astronomy, Vanderbilt University, Nashville, TN 37235, USA}
\newcommand{\AddrAve}{Departamento de F\'{\i}sica da Universidade de Aveiro and CIDMA,  Campus de Santiago, 3810-183 Aveiro, Portugal}

\begin{document}

\title{Stimulated axion decay in superradiant clouds around primordial black holes}

\author{Jo\~{a}o G.~Rosa} \email{joao.rosa@ua.pt}\affiliation{\AddrAve}
\author{Thomas W. Kephart}    \email{thomas.w.kephart@vanderbilt.edu}\affiliation{\AddrVandy}

\date{\today}

\begin{abstract}
The superradiant instability can lead to the generation of extremely dense axion clouds around rotating black holes. We show that, despite the long lifetime of the QCD axion with respect to spontaneous decay into photon pairs, stimulated decay becomes significant above a minimum axion density and leads to extremely bright lasers. The lasing threshold can be attained for axion masses $\mu \gtrsim 10^{-8}\ \mathrm{eV}$, which implies superradiant instabilities around spinning primordial black holes with mass $\lesssim 0.01M_\odot$. Although the latter are expected to be non-rotating at formation, a population of spinning black holes may result from subsequent mergers. We further show that lasing can be quenched by Schwinger pair production, which produces a critical electron-positron plasma within the axion cloud. Lasing can nevertheless restart once annihilation lowers the plasma density sufficiently, resulting in multiple laser bursts that repeat until the black hole spins down sufficiently to quench the superradiant instability. In particular, axions with a mass $\sim 10^{-5}\ \mathrm{eV}$ and primordial black holes with mass $\sim 10^{24}$ kg, which may account for all the dark matter in the Universe, lead to millisecond-bursts in the GHz radio-frequency range, with peak luminosities $\sim 10^{42}$ erg/s, suggesting a possible link to the observed fast radio bursts.
\end{abstract}


\maketitle


It is well-known that a spinning black hole (BH) suffers from the superradiant instability, where light bosonic particles are copiously produced in quasi-bound states around the BH, by extracting its rotational energy \cite{Damour:1976, Zouros:1979iw, Detweiler:1980uk, Furuhashi:2004jk, Cardoso:2005vk, Dolan:2007mj, Konoplya:2008hj, Rosa:2009ei, Konoplya:2011it, Konoplya:2011qq, Dolan:2012yt, review}. This process has attracted a lot of interest as an astrophysical probe of beyond the Standard Model particles, in particular axions \cite{Arvanitaki-JMR, Arvanitaki:2010sy}, as well as higher-spin particles \cite{Rosa:2011my, Pani:2012bp, Brito:2013wya, East:2017ovw, Baryakhtar:2017ngi, East:2017mrj}. All studies have so far neglected the decay of the particles produced by superradiant instabilities, given the long lifetimes expected for light bosons. In particular, the QCD axion \cite{Peccei:1977hh, Weinberg:1977ma, Wilczek:1977pj}, $\phi$, decays into two photons through the interaction \cite{Adler:1969gk, Kaplan:1985dv, Srednicki:1985xd,Sikivie:1988mz}:
\begin{eqnarray} \label{photon_coupling}
\mathcal{L}_{\phi\gamma\gamma} ={\alpha K\over 8\pi F_\phi}\phi F_{\mu\nu}\tilde{F}^{\mu\nu}~,
 \end{eqnarray}
where $\alpha$ is the fine structure constant, $F_{\mu\nu}$ is the Maxwell tensor and $\tilde{F}_{\mu\nu}$ its dual. For an axion with mass $\mu$, its decay constant is $F_\phi \simeq 6\times 10^{11} \left(10^{-5}\ \mathrm{eV}/\mu\right)\ \mathrm{GeV}$ and $K\sim\mathcal{O}(1-10)$ is a model-dependent factor (see e.g.~\cite{Cheng:1995fd}). This yields an axion lifetime:
\begin{eqnarray} \label{axion_lifetime}
\tau_\phi \simeq 3\times 10^{32} K^{-2}\left({\mu \over 10^{-5}\ \mathrm{eV}}\right)^{-5}\ \mathrm{Gyr}~,
\end{eqnarray}
which exceeds the age of the Universe for $\mu \lesssim \mathrm{few}\ \mathrm{eV}$. The QCD axion is, in fact,  a prime dark matter candidate for $10^{-12}\ \mathrm{eV}\lesssim \mu \lesssim 10^{-2}\ \mathrm{eV}$ \cite{Ringwald:2016yge} where its present abundance is set by coherent oscillations \cite{Preskill:1982cy, Abbott:1982af, Dine:1982ah} or the decay of topological defects (see e.g.~\cite{Ballesteros:2016euj}).

However, one must take into account that stimulated decay can become significant in dense environments as the superradiant axion clouds. Stimulated decay has, in fact, been shown to lead to lasing in dense axion clusters when the latter's diameter exceeds the photon's mean free path before it stimulates axion decay \cite{Tkachev, Kephart}. In this Letter, we show, for the first time, that lasing can be triggered in superradiant axion clouds around spinning BHs above a threshold number density set by the axion and BH's properties. Such a threshold can be attained only for spinning sub-stellar mass BHs that result from mergers of non-spinning primordial BHs. This can lead to one of the brightest sources in the cosmos, which we denote as {\it Black hole Lasers powered by Axion SuperradianT instabilities} (BLASTs).

We start by reviewing the superradiant instability for massive scalar fields. In Kerr spacetime, the Klein-Gordon equation $\left(\nabla^\mu\nabla_\mu -\mu^2\right)\phi=0$ admits quasi-bound state solutions, characterized by integer quantum numbers $(n,l,m)$ and localized in the BH's vicinity. In the non-relativistic regime, which is the most relevant to our discussion, one finds a Hydrogen-like spectrum of the form (see e.g.~\cite{review}):
\begin{eqnarray} \label{spectrum}
\hbar \omega_n \simeq \mu c^2\left(1-{\alpha_\mu^2\over 2 n^2}\right)~,
\end{eqnarray}
yielding a ``gravitational atom" where the dimensionless mass coupling is given by:
\begin{eqnarray} \label{alpha_mu}
\alpha_\mu \equiv {G\mu  M_{BH}\over \hbar c} \simeq 0.037\left({\mu\over 10^{-5}\ \mathrm{eV}}\right)\left({M_{BH}\over 10^{24}\ \mathrm{kg}}\right)~,
\end{eqnarray}
where $M_{BH}$ is the BH's mass. These quasi-bound states have complex frequencies $\omega = \omega_R+i\omega_I$, with the sign of the imaginary part determining whether the scalar field grows or decays exponentially as $e^{\omega_I t}$. The superradiant instability corresponds to the former case and occurs whenever $\omega_R <m \Omega_{BH}$, where $\Omega_{BH}$ is the horizon's angular velocity. This instability is powered by the BH's rotational energy, which is extracted by the scalar field and leads to a growing axion cloud around the BH. The fastest growing mode is the ``$2p$" state ($n=2$, $l=m=1$ \cite{foot1}), for which the occupation number grows at a rate:
\begin{eqnarray} \label{superradiant_rate}
\Gamma_s\!\simeq\! {\tilde{a}\over 24}\alpha_\mu^9\!\left(c^3\over GM\right)
\!\simeq\! 4\times 10^{-4}\tilde{a}\!\left({\mu\over 10^{-5}\ \mathrm{eV}}\right)\!\!\left({\alpha_\mu\over 0.03}\right)^8\ \mathrm{s}^{-1}~,\nonumber\\
 \end{eqnarray}
for $\alpha_\mu\ll 1$, where $\tilde{a}= cJ_{BH}/GM_{BH}^2$ is the BH's dimensionless spin parameter ($0<\tilde{a}<1$). 

\begin{figure}[h]\vspace{-0.5cm}
\centering\includegraphics[scale=0.85]{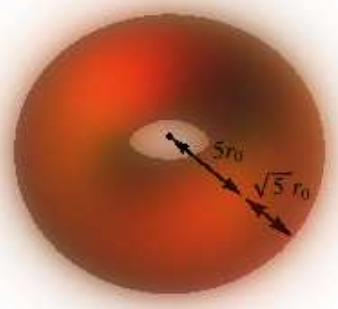}\vspace{-0.5cm}
\caption{Model of the superradiant $2p$-axion cloud around a central Kerr black hole.}
\label{cloud}
\end{figure}

The $2p$-axion cloud has an approximately toroidal shape, as shown in Fig.~\ref{cloud}, with radii $\langle r\rangle = 5r_0$ and $\Delta r= \sqrt{5} r_0$, where the ``Bohr radius" is given by:
\begin{eqnarray}
 r_0= {\hbar\over \mu c \alpha_\mu} \simeq 66 \left({\alpha_\mu\over 0.03}\right)^{-1}\left({\mu\over 10^{-5}\ \mathrm{eV}}\right)^{-1}\ \mathrm{cm}~.
\end{eqnarray}

For comparison, the event horizon is located at:
\begin{eqnarray}
\! r_+\!= \!{GM\over c^2}f(\tilde{a}) \simeq 0.1\! \left({\alpha_\mu\over 0.03}\right)\!\!\left({\mu\over10^{-5}\ \mathrm{eV}}\right)^{-1}\! f(\tilde{a})\ \mathrm{cm}~,
 \end{eqnarray}
where $f(\tilde{a})=1+\sqrt{1-\tilde{a}^2}$, so that in the non-relativistic regime the axion cloud is localized far away from the horizon and we may neglect curvature effects in studying axion decay. Note that the axion's r.m.s.velocity, $\sqrt{\langle v^2\rangle}\simeq(\alpha_\mu/2) c$, is non-relativistic for $\alpha_\mu\ll 1$.

Let us now analyze the dynamics of axion decay within the superradiant cloud. In the axion's rest frame, the two photons must have the same helicity by angular momentum conservation, and we can take this as a first approximation in the non-relativistic regime. The variation of the photon number density for a given helicity $\lambda=\pm$ due to axion decays and inverse decays is given by the Boltzmann equation \cite{Kephart}:
\begin{eqnarray} \label{boltzmann_1}
{dn_\lambda (\mathbf{k})\over dt}&=&\int dX_{LIPS} \big[f_\phi(\mathbf{p})(1+f_\lambda(\mathbf{k}))(1+f_\lambda(\mathbf{k'}))\nonumber\\
&-& f_\lambda(\mathbf{k})f_\lambda(\mathbf{k'})(1+f_\phi(\mathbf{p}))\big]|\mathcal{M}|^2~,
 \end{eqnarray}
where $f_i$ denote the phase space densities of each species, with $n_i= \int d^3k_i/(2\pi)^3 f_i$, $\mathcal{M}$ is the matrix element corresponding to the interaction in Eq.~(\ref{photon_coupling}) and the phase space integration is given by the usual Lorentz-invariant measure including the axion and photon momenta. To integrate this equation, we take the axion and photon phase space distributions as approximately homogeneous and isotropic within the $2p$-cloud, vanishing outside the latter. The maximum axion momentum is then $p_\mathrm{max}\simeq \mu \beta c$, $\beta\equiv \alpha_\mu/2\ll 1$, corresponding to the typical axion momentum within the cloud, and consequently photon momenta are limited by  $\mu c(1\pm\beta)/2$. Although the geometry of the problem is slightly more intricate, this is sufficiently good for the order of magnitude estimates we are mostly interested in. Note that the $2p$-state grows exponentially faster than all others, which may be discarded. This yields for the total photon number density:
\begin{eqnarray} \label{boltzmann_2}
{dn_\gamma\over dt} = \Gamma_\phi\left[2n_\phi\left(1+{8\pi^2\over \mu^3 \beta} n_\gamma\right)- {16\pi^2\over 3\mu^3}\left(\beta+{3\over 2}\right)n_\gamma^2\right]~.
 \end{eqnarray}
where $\Gamma_\phi=\tau_\phi^{-1}$ is the spontaneous axion decay width. In this equation, the first terms within the square brackets correspond to spontaneous and stimulated decay, the latter proportional to $n_\phi n_\gamma$, while the last terms, proportional to $ n_\gamma^2$, correspond to inverse decays (annihilation). The term proportional to $\beta$ corresponds to annihilation into a low-energy ``active" axion, which may remain in the cloud, while the term proportional to the factor $3/2$ corresponds to the production of ``sterile" axions with large velocities, which escape from the system \cite{Kephart}. To obtain the full evolution of the system, we must also take into account the superradiant axion source and that photons escape the cloud at a rate:
\begin{eqnarray} \label{escape_rate}
\Gamma_e = {1\over \sqrt{5}}{c\over r_0} \simeq 2\times 10^{8}\left({\mu\over 10^{-5}\ \mathrm{eV}}\right)\left({\alpha_\mu\over 0.03}\right)\ \mathrm{s}^{-1}~,
 \end{eqnarray}
which is essentially the inverse of the cloud's light-crossing time. This generically yields $\Gamma_\phi \ll \Gamma_s \ll \Gamma_e$.

Thus, we obtain the following system of coupled differential equations for the evolution of the number of axions and photons within the cloud:
\begin{eqnarray} \label{diff_eqs}
{dN_\phi\over dt}& =& \Gamma_s N_\phi -\Gamma_\phi\left[N_\phi(1+AN_\gamma) - B_1N_\gamma^2\right]~,\nonumber\\
{dN_\gamma\over dt} & = & -\Gamma_e N_\gamma + 2\Gamma_\phi\left[N_\phi(1+AN_\gamma) - B N_\gamma^2\right]~,
 \end{eqnarray}
where $A=8\alpha_\mu^2/25$, $B_1= 2\alpha_\mu^4/75$, $B_2= 2\alpha_\mu^3/25$ and $B=B_1+B_2$. 

\begin{figure}[htbp]
\centering\includegraphics[scale=0.87]{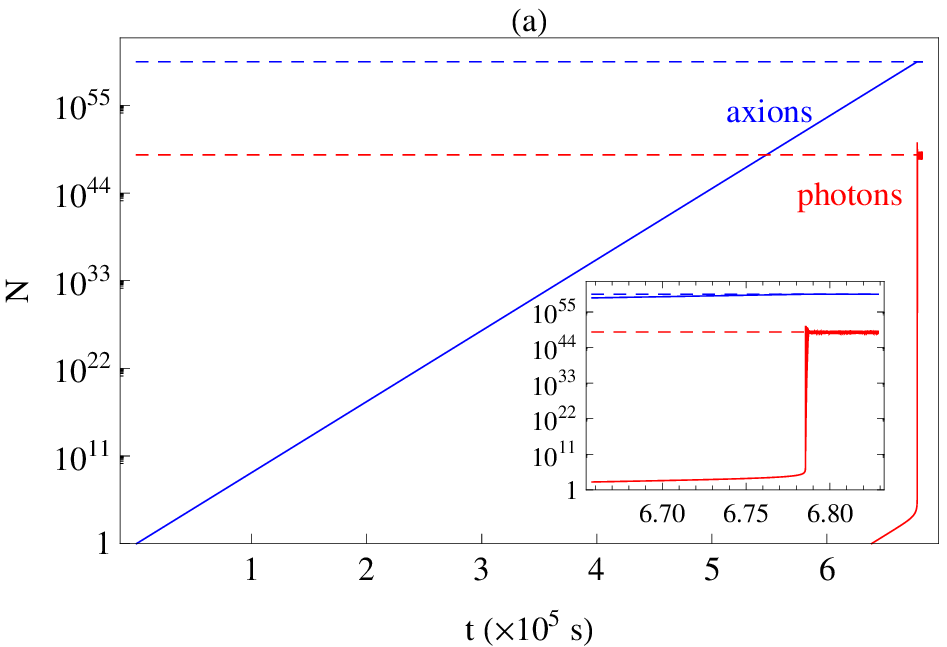}\vspace{0.1cm}
\centering\includegraphics[scale=0.87]{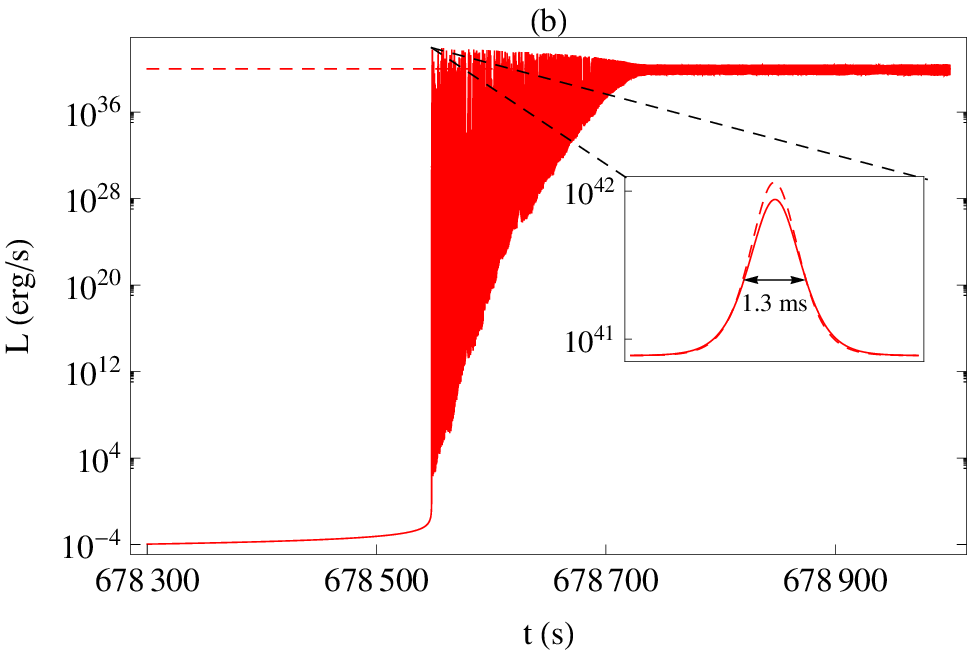}
\caption{Numerical evolution of (a) the axion (blue) and photon (red) numbers within the $2p$-toroidal cloud for an axion with $\mu=10^{-5}\ \mathrm{eV}$ and $K=1$, and a BH with $M_{BH}=8\times 10^{23}$ kg and $\tilde{a}=0.7$ ($\alpha_\mu\simeq 0.03$). The corresponding dashed lines give the critical axion and photon numbers and the inset plot shows a detail of the evolution around the lasing threshold. In (b) we show the evolution of the average photon luminosity after the onset of lasing, with the inset plot zooming into the first laser pulse (solid red curve) and its approximate analytical description (dashed red curve).}
\label{evolution}
\end{figure}

In Fig.~\ref{evolution} we show a numerical solution for a representative choice of parameters.
As clear in this figure, the number of axions starts growing exponentially, $N_\phi\sim  e^{\Gamma_s t}$, due to the superradiant instability. Their spontaneous decay then yields an exponentially growing photon number $N_\gamma\simeq (2\Gamma_\phi/\Gamma_e) N_\phi$, until $N_\gamma\sim A^{-1}$ and stimulated decay becomes significant. This occurs when $N_\phi \simeq N_\phi^c= \Gamma_e/ (2A\Gamma_\phi)$, which corresponds to a stable critical point. Once stimulated decay takes over and lasing begins, the number of photons rises sharply and then exhibits damped oscillations about the equilibrium value $N_\gamma^c= \Gamma_s/(A\Gamma_\phi)$. These oscillations, which effectively constitute ``laser pulses", are the result of stimulated axion decay and photon escape dominating alternately, and their damping is due to photon annihilation. Note that, apart from the particular toroidal geometry of the problem and the fact that lasing occurs in multiple directions within the axion cloud, these laser pulses are otherwise very similar to conventional lasers, peaking at a frequency corresponding to half the axion's mass.

At the onset of lasing, we may neglect annihilation and take $X\!=\!|N_\phi/N_\phi^c-1|\!\ll\!1 $ and $Y\!\!=\!N_\gamma/N_\phi^c\ll 1$, obtaining $d^2\log Y/du^2 \!=\! \eta - Y/2$, where $u\!\!=\!\!\Gamma_e t$ and $\eta=\Gamma_s/\Gamma_e\ll 1$. This equation has solutions of the form $Y_I=Y_0e^{\eta u^2/2}$ for $Y\ll 2\eta$, and $Y_{II} = C\cosh^{-2}\left(\sqrt{C}(u-u_{\mathrm{max}})/2\right)$ for $Y\gg 2\eta$, which can be matched smoothly at $Y= 2\eta$, taking $N_\gamma\sim A^{-1}$ at the onset of lasing. This yields a good approximation to the numerical solution, as clear in Fig.~\ref{evolution}(b), despite a slight overestimation of the luminosity $L=\Gamma_e\mu c^2N_\gamma/2$ due to some uncertainty in $Y_0$. We may then estimate the peak luminosity, total energy and duration (FWHM) of the first laser pulses or ``bursts", which are approximately given by:
\begin{eqnarray}  \label{BLAST_parameters}
L_{B}&\simeq &{2\times 10^{42}\over K^2}\tilde{a} \left({10^{-5}\ \mathrm{eV}\over \mu}\right)^{2}\left(\alpha_\mu\over 0.03\right)^7\left({\xi\over 100}\right)\ \mathrm{erg/s}~,\nonumber\\
E_{B}&\simeq& {3\times 10^{39}\over K^2}\sqrt{\tilde{a}} \left({10^{-5}\ \mathrm{eV}\over \mu}\right)^{3}\left(\alpha_\mu\over 0.03\right)^{5\over2}\left({\xi\over 100}\right)^{1\over2}\ \mathrm{erg}~,\nonumber\\
 \tau_{B}&\simeq &{1\over \sqrt{\tilde{a}}} \left({10^{-5}\ \mathrm{eV}\over \mu}\right)\left(\alpha_\mu\over 0.03\right)^{-9/2}\left({\xi\over 100}\right)^{-1/2}\ \mathrm{ms}~,
  \end{eqnarray}
where $\xi= \log\left(\Gamma_s/\Gamma_\phi\right)\simeq  107-4\log\left(\mu/ 10^{-5}\ \mathrm{eV}\right)+8\log\left(\alpha_\mu/ 0.03\right)+\log \left({\tilde{a}/ K^2}\right)$.
%
%
Note that the peak luminosity exceeds the equilibrium value by $\xi\sim \mathcal{O}(100)$.

We thus find that BLASTs can have extremely large peak luminosities, even for $\alpha_\mu\ll 1$ and moderate BH spins. In fact, the brightest BLASTs will generically exceed the BH's Eddington luminosity, $L_{Edd}\simeq 10^{38}\left(M/ M_\odot\right)~\mathrm{erg/s}$, so that any interstellar plasma surrounding the BH should be blown away by the radiation pressure. However, the intense electromagnetic field of the laser may be close to the Schwinger limit for $e^+e^-$ pair production \cite{Sauter:1931zz, Schwinger:1951nm}. Although for a single plane wave laser beam Schwinger pair production does not occur since $ F_{\mu\nu}F^{\mu\nu}=F_{\mu\nu}\tilde{F}^{\mu\nu}=0$, lasing occurs in multiple intersecting directions within the axion cloud and these invariants should be non-trivial (see e.g.~\cite{Ringwald:2001ib}). The electric field created by the BLAST within the cloud is approximately given by \cite{foot2}: 
\begin{eqnarray} 
|\mathbf{E}| \sim  E_c \left(\mu\over 10^{-5}\ \mathrm{eV}\right)\left(\alpha_\mu\over 0.03\right)\left({L\over 10^{43}\ \mathrm{erg/s}}\right)^{1/2}~,
 \end{eqnarray}
where $E_c\simeq 1.3\times 10^{18}\ \mathrm{Vm}^{-1}$ is the critical Schwinger field. Since pair production may be significant just below this critical value, we expect the brightest BLASTs to generate a dense $e^+e^-$ plasma within the cloud. This plasma may then quickly reach the critical density for photon propagation, $n_c\sim 10^{12} (\mu/ 10^{-5}\ \mathrm{eV})^2\ \mathrm{cm}^{-3}$, at which point the photon plasma mass blocks axion decay and lasing stops. Lasing may thus end after a single laser pulse and at least the brightest BLASTs should thus correspond to short radiation bursts with extremely high luminosities, yielding true ``black hole bombs" \cite{Press:1972zz}.  

After lasing stops the non-relativistic plasma is confined by the BH,  but its density eventually decreases due to $e^+e^-$ annihilations, with a lifetime $\tau_{ann}\simeq 4\left(n_e/ 10^{12}\ \mathrm{cm^{-3}}\right)^{-1}\ \mathrm{hours}$ for direct annihilation, which can be reduced by positronium formation for low temperatures, and possibly other environmental effects \cite{Greaves}. Lasing may then restart once the plasma becomes sub-critical, leading to repeating laser bursts. 

Since lasing requires a threshold axion number $N_\phi^c$, we must require that the threshold cloud's mass and spin are smaller than the corresponding BH parameters, or otherwise the superradiant instability may shut down before lasing commences. The strongest constraint is on the spin of the critical axion cloud, $J_\phi^{c}=\hbar N_\phi^c$, yielding:
\begin{eqnarray} \label{spin_bound}
{J_\phi^{c}\over J_{BH}}\!=\!{{1\over \tilde{a}\alpha_\mu}{M_\phi^c\over M_{BH}}}\!\simeq\! {0.06\over \tilde{a}\alpha_\mu^3 K^2}\left({10^{-8}\,\mathrm{eV}\over\mu}\right)^{2} \lesssim  1.
 \end{eqnarray}

Since $\alpha_\mu\lesssim 0.5$ for superradiant modes, lasing can only occur for axions with $\mu\gtrsim 10^{-8}$ eV and consequently BHs with $M_{BH}\lesssim 10^{-2}M_\odot$. 

Non-linear axion self-interactions may also prevent the cloud from reaching the lasing threshold, since they mix superradiant and non-superradiant states. This results in the BH partially absorbing the cloud in a ``bosenova" collapse, which keeps the axion field $\phi\lesssim F_\phi$ \cite{Arvanitaki-JMR, Arvanitaki:2010sy, Yoshino:2012kn}. Since $\rho_\phi\simeq \mu n_\phi \sim \mu^2\phi^2$, we can estimate the number of axions within the cloud for which quartic self-interactions with coupling $\lambda\simeq 0.3 \mu^2/F_\phi^2$ \cite{diCortona:2015ldu} become significant, and require this to exceed $N_\phi^c$. This implies:
\begin{eqnarray} \label{alpha_bound}
\alpha_\mu\lesssim 0.03 K~,
\end{eqnarray}
so that for $K\sim 1$ lasing can only occur in the non-relativistic regime, as anticipated \cite{foot3} (although $K\sim \mathcal{O}(10)$ in some axion models \cite{Cheng:1995fd}). 

We thus find that BLASTs cannot occur for BHs resulting from stellar collapse, being nevertheless possible for primordial BHs (PBHs) \cite{Carr:1974nx, Carr:1975qj, Carr:1976zz, Meszaros:1975ef, GarciaBellido:1996qt, Khlopov:2008qy} that may have formed in the early Universe as the result of large density fluctuations in,  e.g.,~some inflationary scenarios \cite{Lyth:2011kj, Bugaev:2011wy, Clesse:2015wea, Garcia-Bellido:2017mdw, Carr:2017edp, Germani:2017bcs, Ballesteros:2017fsr}, curvaton models \cite{Kohri:2012yw, Kawasaki:2012wr, Bugaev:2013vba} or cosmological phase transitions \cite{Jedamzik:1999am}. PBHs may last until the present day for $M_{BH}\gtrsim 10^{12}$ kg and significantly contribute to the dark matter abundance in the $(10^{-10} - 10^{-4})M_\odot$ mass range   \cite{Carr:2009jm, Frampton:2010sw, Pani:2013hpa, Belotsky:2014kca, Kuhnel:2017pwq, Carr:2017jsz}. 

PBHs are born with no spin, since they result from the collapse of overdensities in an isotropic gas \cite{Garcia-Bellido:2017fdg}, and are unlikely to spin up due to accretion, which occurs in a nearly spherical configuration for sub-stellar masses \cite{Ali-Haimoud:2016mbv}. They may, however, merge into BHs with larger mass and spin, triggering superradiant instabilities, with e.g.~the merger of two non-spinning BHs of similar masses yielding a final spin $\tilde{a}_f\simeq 0.7$ \cite{Scheel:2008rj}. If PBHs are mostly clustered in high dark matter density regions, the expected rate at which PBHs are captured in binary systems that subsequently merge is given by \cite{Clesse:2016vqa}:
\begin{eqnarray} \label{merger}
\Gamma_{\mathrm{capt}}^{\mathrm{total}}\simeq 3\times 10^{-9}f_{\mathrm{DM}}\delta^{\mathrm{loc}} \ \mathrm{yr}^{-1}\mathrm{Gpc}^{-3}~,
\end{eqnarray}
where $f_{DM}$ is the dark matter fraction in PBHs of the relevant mass and $\delta^{\mathrm{loc}}$ is the local density enhancement due to clustering \cite{foot4}. The latter may reach values $\sim10^9-10^{10}$ in globular clusters and in dark matter-dominated ultra-faint dwarf galaxies. We may thus expect up to a few new BLASTs to be triggered across the observable Universe every year for $f_{DM}\lesssim 0.1$.

A particularly interesting case is that of a QCD axion with $\mu\simeq 10^{-5}$ eV, which may be the dominant dark matter component. For $K\sim 1$, lasing may occur for PBHs with $M_{BH}\sim (2-8)\times 10^{23}$ kg ($\alpha_\mu\simeq 0.01-0.03$) according to Eqs.~(\ref{spin_bound}), (\ref{alpha_bound}) and the superradiance condition. PBHs in this mass range could account for up to 12$-$24\% of dark matter according to microlensing surveys \cite{Tisserand:2006zx}, and generate BLASTs in the GHz range with peak luminosity $L_B\sim 10^{39}-10^{42}\ \mathrm{erg/s}$ and pulse duration $\tau_B\sim 1-100$ ms. 

On the one hand, the less luminous BLASTs, associated with the lighter BHs, are below the Schwinger limit. They may then reach the equilibrium configuration and yield continuous laser sources lasting for $\mathcal{O}(1-100)$ years until the BH's rotational ``fuel" is exhausted. The brightest BLASTs, on the other hand, should yield multiple millisecond bursts with peak luminosities $\sim 10^{42}\ \mathrm{erg/s}$, thus exhibiting tantalizing similarities with the several fast radio bursts (FRBs) observed in recent years \cite{Lorimer:2007qn, Thornton:2013iua, Petroff:2016tcr}, in particular FRB 121102  has been shown to repeat and is localized within a faint dwarf galaxy \cite{Chatterjee:2017dqg, Marcote:2017wan, Tendulkar:2017vuq}. (See \cite{Conlon:2017hhi} and \cite{Tkachev:2014dpa, Iwazaki:2017rtb} for alternative FRB models involving BH superradiance or axion miniclusters, respectively. For further discussion of FRBs see \cite{Kehayias:2015vba} and references therein.)

A rigorous comparison between BLASTs and FRBs requires a more detailed analysis, taking into account geometrical and curved spacetime effects that we have neglected for simplicity. In particular, lasing should occur mostly in the equatorial plane, where photons can travel larger distances within the cloud and stimulate more axion decays. This could prevent the observation of some bursts along our line of sight, depending on the BH's spin axis and trajectory, and possibly explain why some FRBs do not repeat. Large enhancements in the local plasma density or temperature, which induce a significant photon mass, may also temporarily block lasing. 

Scattering of laser photons by the BH may also produce interesting, albeit sub-leading, effects. In particular, by angular momentum conservation one of the photons from each axion decay co-rotates with the black hole and satisfies the superradiance condition $\omega_R< m\Omega_{BH}$, while the other is in a non-superradiant state. This may slightly decrease the laser luminosity and, more interestingly, modify its polarization through the spin-helicity effect (see e.g.~\cite{Rosa:2015hoa, Leite:2017zyb}). A more detailed analysis of laser-plasma interactions and Schwinger pair production is also required, since these could potentially induce both spectral and temporal distortions of the bursts and influence their repetition pattern. Axion-photon conversion in a magnetic field could also, in principle, modify the BLAST dynamics, but plays a negligible role for the small ambient magnetic fields $\sim \mu\mathrm{G}$ estimated for dark matter-dominated regions such as faint dwarf galaxies. Nevertheless, if dark matter is mostly made up of $10^{-5}$ eV axions, their conversion into photons in the magnetic field of our galaxy could be detected by the Square Kilometer Array (SKA) \cite{Kelley:2017vaa}.

Our estimates nevertheless clearly suggest a possible link between the brightest GHz BLASTs and FRBs. For the brightest BLASTs, each BH may generate up to $\sim 10^8$ bursts before the superradiant instability shuts down, if energy loss through sterile axions and gravitational waves can be neglected, remaining active for up to $\sim 10^4$ years. This yields up to $\sim 10^5$ FRBs per day across the whole sky. In addition, one may expect a companion population of continuous laser sources at the same frequency but with lower luminosities, associated with the lighter primordial BHs in the mass range given above. These should be less numerous than repeating BLASTs due to their considerably shorter lifetime.

Potential signatures of the BLAST nature of FRBs could be associated with $e^+e^-$ annihilation and/or positronium afterglows, or gravitational wave bursts from bosenova collapse in between bursts (noting that the axion number grows while lasing is blocked), to be explored in future work. A clear prediction is the existence of an axion with a mass  $\sim 10^{-5}$ eV and coupling parameter $K\sim 1$. There are already ongoing experiments searching for axion dark matter in this mass range, including ADMX \cite{Stern:2016bbw}, X3 \cite{Brubaker:2016ktl} and CULTASK \cite{Petrakou:2017epq}, and planned experiments such as MADMAX \cite{TheMADMAXWorkingGroup:2016hpc} and ORPHEUS \cite{Rybka:2014cya}. These will be complemented with searches in different mass ranges (e.g.~CASPEr \cite{Budker:2013hfa}, ABRACADABRA \cite{Kahn:2016aff}, QUAX \cite{Barbieri:2016vwg}), fifth force experiments such as ARIADNE \cite{Geraci:2017bmq} and helioscopes  such as IAXO \cite{Armengaud:2014gea}. We note that the occurrence of BLASTs is independent of the QCD axion accounting for dark matter and, as such, can also be probed with e.g.~light-shining-through-a-wall experiments such as OSQAR \cite{Ballou:2015cka} and ALPS II \cite{Bahre:2013ywa}.

FRB detections should also increase dramatically in the coming years, with the advent of telescopes such as CHIME \cite{Amiri:2017qtx}, Apertif \cite{Maan:2017uts} and particularly the SKA \cite{Macquart:2015lsa}. This will lead to a much better FRB characterization and localization, helping to determine whether BLASTs can account for at least a fraction of these bright radio transients. We note that several less exotic FRB progenitors have been proposed in the literature, including merging or collapsing compact objects, or energetic young pulsars and magnetars (see e.g.~\cite{Rane:2017}). We also hope that our results motivate future microlensing surveys to further constrain the PBH abundance in the $10^{-7}-10^{-6}M_\odot$ mass range required for GHz BLASTs, possibly with DECam or LSST along the lines proposed in \cite{Battaglieri:2017aum} for intermediate mass BHs. We emphasize that BLASTs require rather exotic spinning PBHs, which may result from mergers within the original population, since merging Schwarzschild BHs invariably lead to Kerr BHs.

We note that similar lasing events may be induced by other bosonic particles decaying into photons, potentially providing additional signatures for new physics. From the model building perspective, our work also motivates investigating inflationary scenarios leading to mixed PBH-axion dark matter (see e.g.~\cite{Inomata:2017uaw}). In future, we also plan to explore whether the cosmic radio background photons may stimulate axion decay in superradiant clouds and thus help to trigger BLASTs. Finally, it may prove fruitful for more detailed calculations to study BLASTs directly in terms of the coupled axion and Maxwell field equations in the Kerr spacetime.

In conclusion, we have shown that, in conjunction, two of the ``darkest" dark matter candidates, axions and (spinning) PBHs, may lead to some of the brightest sources in the Universe.

\vfill

\vspace{0.2cm} 
\begin{acknowledgments}

We would like to thank Jorge Vieira for useful discussions. J.\,G.\,R. is supported by the FCT Investigator Grant No.~IF/01597/2015 and partially by the H2020-MSCA-RISE-2015 Grant No. StronGrHEP-690904 and by the CIDMA Project No.~UID/MAT/04106/2013.
The work of T.~W.~K.was supported in part by US DoE grant DE-SC0010504.

\end{acknowledgments}


\begin{thebibliography}{99}
\bibitem{foot1} Since the axion is a real scalar field there is also an analogous $l=1$, $m=-1$ superradiant bound state growing at the same rate as $e^{i\omega t}$, but this will not affect our discussion.
\bibitem{foot2} We neglect the magnetic field for simplicity, but this does not change the magnitude of the effect for $|\mathbf{E}| \sim |\mathbf{B}|$.
\bibitem{foot4}
We note that there is a substantial theoretical uncertainty in the distribution of PBHs that could significantly modify the associated merger rates.
\end{thebibliography}
\end{document}